# The Allen Telescope Array: The First Widefield, Panchromatic, Snapshot Radio Camera

**Don Backer, Amber Bauermeister, Leo Blitz, Douglas Bock, Geoffrey C. Bower, Calvin Cheng, Steve Croft, Matt Dexter, Greg Engargiola, Ed Fields, Rick Forster, Colby Gutierrez-Kraybill, Carl Heiles, Tamara Helfer, Susan Jorgensen, Garrett Keating, Casey Law, Joeri van Leeuwen [1], John Lugten, Dave MacMahon, Oren Milgrome, Douglas Thornton, Lynn Urry, Jack Welch, Dan Werthimer, Peter Williams, Melvin Wright**
*Radio Astronomy Laboratory, University of California*
*Berkeley CA, USA*

**Robert Ackermann, Shannon Atkinson, Peter Backus, William Barott, Tucker Bradford, Michael Davis, Dave DeBoer, John Dreher, Gerry Harp, Jane Jordan, Tom Kilsdonk, Tom Pierson, Karen Randall, John Ross, Seth Shostak, Jill Tarter**
*SETI Institute*
*Mountain View CA, USA*

The first 42 elements of the Allen Telescope Array (ATA-42) are beginning to deliver data at the Hat Creek Radio Observatory in Northern California. Scientists and engineers are actively exploiting all of the flexibility designed into this innovative instrument for simultaneously conducting panoramic surveys of the astrophysical sky. The fundamental scientific program of this new telescope is varied and exciting; some of the first astronomical results will be discussed.

---

[1]   *Speaker*
     E-mail: `leeuwen@astron.nl`

# 1. Introduction

The Allen Telescope Array (ATA) is a "Large Number of Small Dishes" (LNSD) array designed to be sensitive for commensal surveys of conventional radio astronomy projects and SETI targets at centimeter wavelengths. It is well known [1] that for surveys requiring multiple pointings of the array antennas to cover a large solid angle in a fixed amount of time, the resulting point source sensitivity is proportional to ND, where N is the number of dishes, and D is the dish diameter, rather than $ND^2$, the total collecting area. Reasonable expectations for antenna and electronics costs then lead to the LNSD array as the optimum.

The ATA will consist of 350 6m-diameter dishes when completed, which will provide an outstanding survey speed and sensitivity. In addition, the many antennas and baseline pairs provide a rich sampling of the interferometer uv plane, so that a single pointing snapshot of the array of 350 antennas yields an image in a single field with about 15,000 independent pixels. Other important features of the ATA include continuous frequency coverage over 0.5 GHz to 10 GHz and four simultaneously available 600-MHz bands at the back-end which can be tuned to different frequencies in the overall band.

The ATA is a joint project of the Radio Astronomy Laboratory of the University of California, Berkeley and the SETI Institute in Mountain View, CA. The initial design grew out of planning meetings at the SETI Institute summarized in the volume "SETI 2020" [2]. The design goals were (a) continuous frequency coverage over as wide a band as possible in the range 1 – 10 GHz, (b) an array cost improvement approaching a factor of 10 over current array construction practices, (c) large sky coverage for surveys, (d) a collecting area as large as one hectare for a point source sensitivity competitive with other instruments, (e) interference mitigation capability for both satellite and ground based interference sources, and (f) both imaging correlator and beamformer capability with rapid data reduction facilities.

The ATA is now complete to 42 antennas. Highlights of the system are the frequency agility, the low background and sidelobes of the antennas, the wideband feed and input receiver, the analog fiber optical system, the large spatial dynamic range, the backend processing systems and the overall low cost.

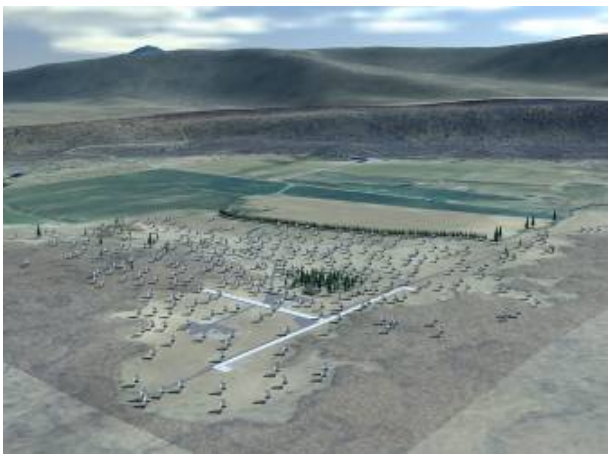

Fig. 1: Rendering of final ATA configuration with 350 antennas and at the Hat Creek Radio Astronomy Observatory (HCRO)

## 2. The Antennas

The antenna is an offset Gregorian design that allows a larger secondary with no aperture blockage for good low frequency performance and also provides a clear aperture with lower sidelobes in the antenna pattern and lower thermal background. The primary is an approximately 6m diameter section of a paraboloid, and the secondary is a 2.4m ellipsoid, four wavelengths across at the lowest operating frequency.

The hydroforming technology used to make these surfaces is the same technique used to generate low-cost satellite reflectors. The pyramidal feed is protected from the weather by a radio transparent radome cover as well as by the shroud. Sidelobes of the mirror and feed system fall largely on the sky, and ground spillover is small. Radio holographic measurements of the entire antennas show overall optical surfaces with total RMS errors of 0.7mm for night time observations. In the middle of the day sunshine doubles the error to about 1.5 mm rms, 1/20 wavelength at our shortest operating wavelength.

## 3. The Feeds and receivers

The ATA feed is a pyramidal log-periodic feed [3]. The motivation for the choice of this wideband feed was the development of very wideband low noise MMIC receivers [4]. The novel feed allows low-noise amplifiers to be housed close to the feed end, thus minimising cable losses and receiver noise. The linear dimensions of the feed yield an operating range from about 500 MHz to 10 GHz, and a linear drive allows the feed focus to be accurately set at the focal point of the two mirror system at any frequency. The overall aperture efficiency with the feed in focus averages 60% over the whole band.

Figure 2 shows total $T_{sys}$ measurements for one of the receivers based on observations of the moon at the higher frequencies and Cas A for the lower frequencies. A constant 60% aperture efficiency was assumed. The agreement with the predictions, ~40K up to 5 GHz, is good, but the temperatures rise up to 80K above 8.5 GHz at higher frequencies. It appears that the input losses are higher than predicted at the higher frequencies. The results are quite good, in any case.

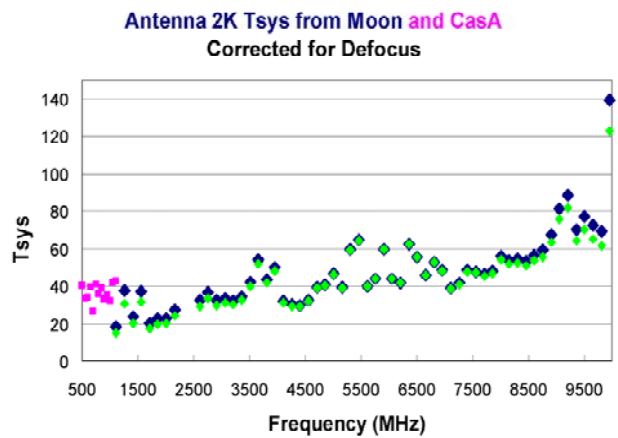

Fig. 2: Measured system temperature for antenna 2K using the moon and Cassiopeia A for reference, assuming 60% efficiency.

**4. The Array**

Figure 3 is an aerial photograph of the present array of 42 antennas. The construction tent is evident in the upper left of the picture, and the processor building is near the upper center of the picture. Figure 4a shows a snapshot distribution of baselines for the 42-antenna array. Figure 4b shows a snapshot of a 98-element array of baselines which is a possible next stage for the growth to the final 350. Figure 4c shows the baseline distributions for the final ATA-350. The azimuthal distribution of baseline angular directions for the ATA-350 is uniform, while the distribution of baseline lengths approaches a Gaussian as closely as possible for the ATA-350 (see Figure 4d). The final configuration of the ATA-350 is a nearly perfect Gaussian in the uv coordinates, except for the zero spacing. The Fourier transform of this distribution is the beam pattern, which is therefore also a Gaussian and an ideal point spread function.

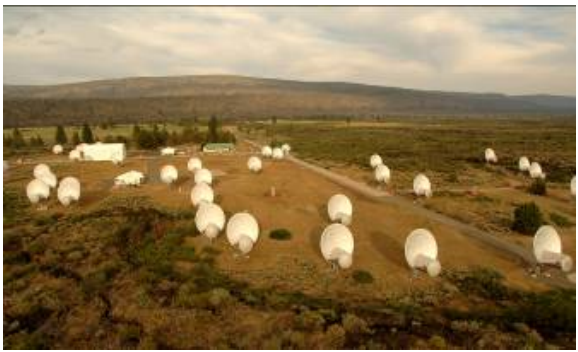

Fig. 3: Aerial view of the ATA-42 at the Hat Creek Radio Astronomy Observatory in Northern CA.

**5. The Analog Electronic System**

The remainder of the signal path lies within the processor building, which is near the center of the eventual 350-element array (see Figure 1). The full 500 MHz to 10 GHz band of each polarization of each antenna is brought back to the processor building over single mode optical fiber as an analog signal [5]. This signal is converted to baseband in an RF converter board (RFCB). Output of four mixers, each fed by a different local oscillator, passes through 600 MHz wide passive bandpass filters and next through 200 MHz wide anti-aliasing filters. Because the four different first LOs are tuned separately, it is possible to perform simultaneous observations at four different frequencies.

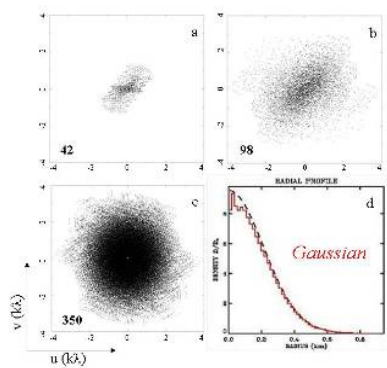

Fig. 4: Snapshot of baseline distributions at 1.42 GHz and 10° declination for a) 42 antennas, b) 98 antennas, and c) 350 antennas. Panel d) is the distribution of baseline lengths for 350 antennas.

## 6. Digital Signal Processors

The great flexibility of the ATA is enabled by multiple signal processing backends that can work simultaneously on the same, or different, independently tunable IF channels. At present the digital backends are spectral-imaging correlators, fast spectrometers for individual antennas, and beamformers feeding SETI signal processors and pulsar processors. More backends are expected in the future to accommodate different types of science.

*A. Digital Correlators:* The ATA correlator relies on an "FX" architecture that first divides the signal into frequency channels then cross-multiplies the result for each antenna pair [6]. Bit-selection in each of the 1024 spectral channels permits a large dynamic range. The present correlators have an overall bandwidth of 100 MHz with 1024 spectral channels. The overall bandwidth can be made smaller in steps of two to increase spectral resolution. The 250-antenna array will produce snapshot maps with 250 independent pixels over the primary beam, but that number can be increased easily by a factor of a few using the traditional earth rotation synthesis.

*B. Beamformers:* The other basic data processing scheme is the construction of beamformers. For this we simply add the outputs of the antennas phased up for a particular direction in the sky over the 100 MHz bandwidth.

## 7. Early science results

*HI in Galaxy Groups*

With only the current reservoir of molecular gas, galactic star formation will cease much sooner than the decline of the observed star formation rate suggests. This is the molecular gas depletion problem. One aspect of the solution to this problem is including atomic hydrogen in the reservoir of gas for star formation. However, to form stars, the gas must be transported to the inner regions of galaxy disks: through inflow of HI from the outer disk or infall of intergalactic HI. We propose a mechanism for this is angular momentum loss through weak tidal interactions between galaxy group members. To constrain the contribution from this effect, we are imaging galaxy groups in the Local Volume (within 10Mpc) in the 21cm line with the ATA. This survey will look for extensions of the HI disks as well as intergalactic HI [7].

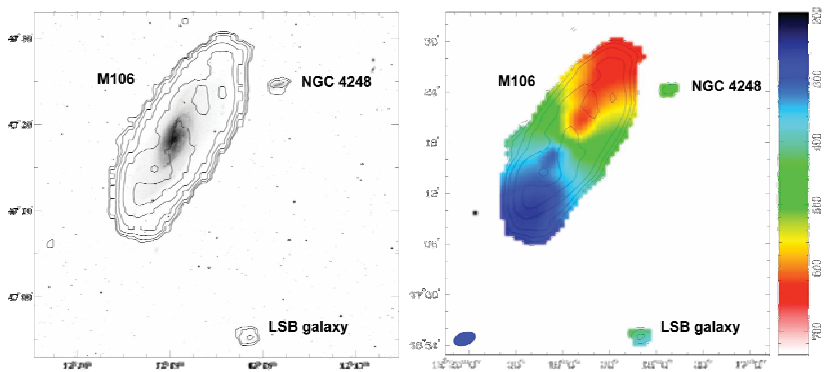

Fig. 5: M106. Moment 0 contours overlaid on DSS image (left) and moment 1 map with velocity colors in km/s (right).

With its small dishes, wide field imaging is a strength of the ATA. In Fig. 10 we show the detection of HI gas in the inter-galactic medium of the Leo group. This 1 square degree region is 20% of the full ATA field of view. GALEX observations have recently revealed ongoing star formation in this cloud, raising the question of how star formation proceeds in a zero metallicity environment that is analogous to the earliest epochs of star formation.

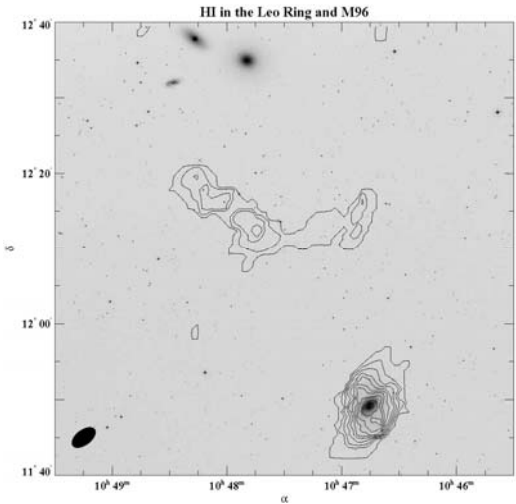

Fig. 6: HI gas in the inter-galactic medium of the Leo group.

*The ATA Twenty-centimeter Survey (ATATS)*

We have recently completed the ATATS of 800 square degrees of sky [8]. We are able to construct light curves to look for changes from epoch to epoch (including transient sources appearing in a single epoch) as well as a deep, multi-epoch image and associated catalog of the sky, which compare well to previous surveys such as NVSS. We use a pipeline which allows us to reduce and analyze the data with a minimum of human interaction.

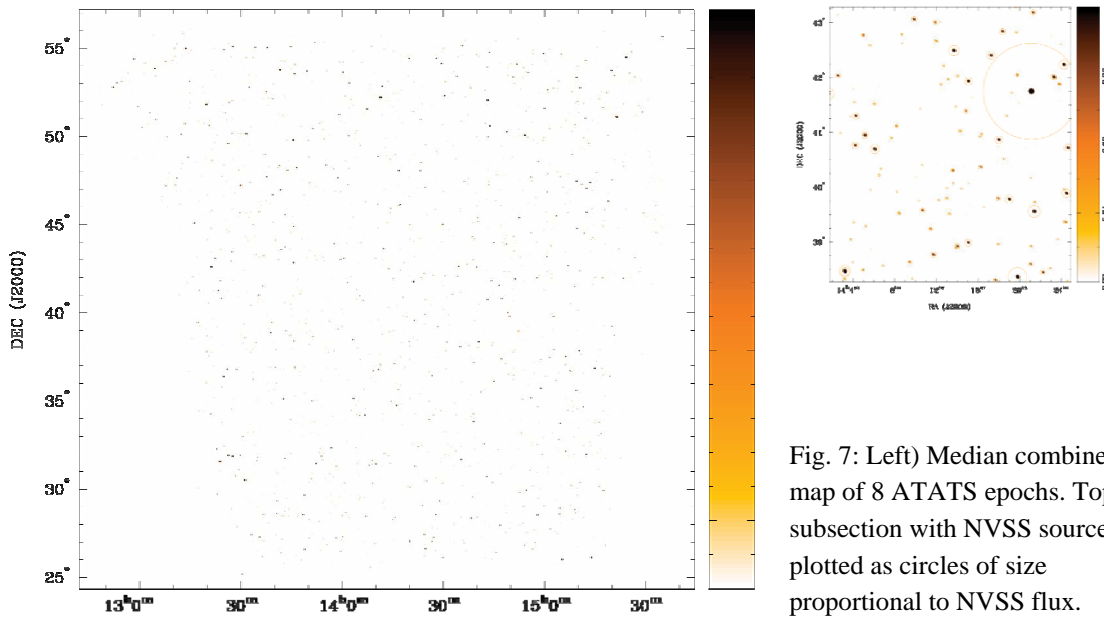

Fig. 7: Left) Median combined map of 8 ATATS epochs. Top) subsection with NVSS sources plotted as circles of size proportional to NVSS flux.

*The Pi GHz Sky Survey (PiGGS)*

The principle goal of PiGGS is exploration of the static and transient radio sky at 3.1GHz, at flux densities matching FIRST and NVSS. We cover a large fraction of the Northern sky and explore time scales from days to months through a tiered approach. Specific results of the survey will be detection of 250,000 radio sources in a 10,000 square degree region of the North Galactic Cap; daily monitoring of a 10 square degree region; and automated real-time identification of transient sources [9].

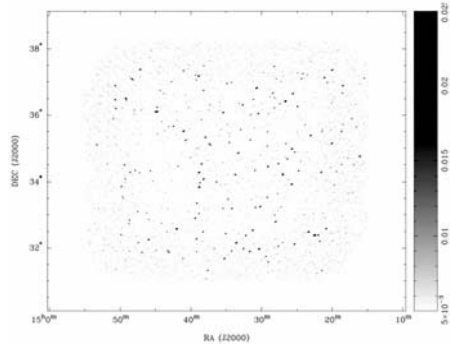

Fig. 8: The Bootes Field, observed daily in PiGGS. 3.1GHz is a sweet spot in terms of field of view, array performance and low RFI for the ATA and enables us to probe spectral indices.

*Continuous Spectra of Star-Forming Galaxies*

Broadband multi-frequency spectra of M82 show the necessity of covering a wide frequency range for characterizing the radio continuum emission from star-forming galaxies and AGN. The spectrum curvature provides insight into the cooling processes in the galaxy [10].

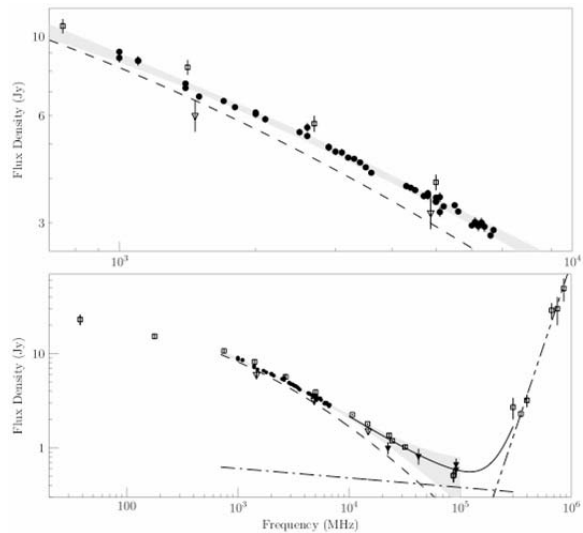

Fig. 9: Wide-band spectra of M83. Top) 700MHz – 10GHz. Bottom) 40 MHz – 800 GHz.

| *Number of Elements, N* | *42* | *256* | | |
|---|---|---|---|---|
| Element Diameter | 6.1 | 6.1 | m | projected on sky |
| Total Geometric Area | 1227 | 7478 | m2 | |
| Aperture Efficiency | 63% | 63% | | |
| System Temperature | 44 | 44 | K | average across 1-10 GHz |
| SEFD | 156 | 26 | Jy | |
| Array Diameter | 0.3 | 0.8 | km | |
| Field of View | $3.5/f_{GHz}$ | $3.5/f_{GHz}$ | degrees | |
| Antenna Pointing | 0.01 | 0.01 | beam | 10 GHz, night, low winds |
| Slew time | 2 | 2 | minutes | to anywhere on sky |
| Frequency Coverage | 0.5-11.2 | 0.5-11.2 | GHz | available simultaneously |
| Synthesized Beam | Gaussian | Gaussian | | |
| Max. Deviation from Gaussian | 18% | 1% | | "snapshot" |
| FWHM of Synthesized Beam | 248x120 | 93x76 | $arcsec^2$ | 1.4 GHz @ zenith |
| Sensitivity | | | | |
| Beamformer Bandwidth | 3 x 72 | 8 x 500 | MHz | |
| Correlator Bandwidth | 2 x 100 | 2 x 500 | MHz | Reducible by $2^N$ |
| Correlator Channels | 2 x 1024 | 2 x 2048 | | |
| Continuum Point Source | 590 | 43 | uJy | rms in 6 min |
| Brightness Temperature | 34 | 24 | mK | rms in 12h @ 10 km/s in HI |
| Continuum Survey Speed | 0.039 | 7.24 | $deg^2\ s^{-1}$ | 1 mJy rms @1.4 GHz |
| $T_B$ Survey Speed | 0.096 | 0.202 | $deg^2\ s^{-1}$ | 1K rms in 10 km/s @1.4 GHz |

Table 1: Parameters for ATA-42 and ATA-256